\documentclass{eas}
\usepackage{graphicx}
\newcommand{\hmpc}{h^{-1}{\rm Mpc}}

\newcommand{\gad}{{\sc Gadget-2}}
\newcommand{\ion}[2]{\hbox{#1\,{\sc #2}}}
\begin{document}

\title{The Mass-Metallicity Relation in Cosmological Hydrodynamic Simulations} 
\runningtitle{Simulations of the Mass-Metallicity Relation}
\author{Romeel Dav\'e$^1$}
\author{Kristian Finlator$^1$}
\author{Benjamin D. Oppenheimer}\address{Steward Observatory, 933 N. Cherry Ave., Tucson, AZ, USA}

 \begin{abstract}
We use cosmological hydrodynamic simulations with enriched galactic
outflows to compare predictions for the galaxy mass-metallicity ($M_*-Z$)
with observations at $z\approx 2$ from \cite{erb06}.  With no outflows
included galaxies are over-enriched, indicating that outflows are required
not only to suppress star formation and enrich the IGM but also to lower
galaxy metal content.  The observed $M_*-Z$ slope is matched both in
our model without winds as well as our favored outflow model where the
outflow velocity scales as the escape velocity, but is too steep in
a model with constant outflow speeds.  If outflows are too widespread
at early times, the IGM out of which smaller galaxies form can become
pre-polluted, resulting in a low-mass flattening of the $M_*-Z$ relation
that is inconsistent with data.  Remarkably, the same momentum-driven
wind model that provides the best agreement with IGM enrichment data
also yields the best agreement with the $z\approx 2$ $M_*-Z$ relation,
showing the proper outflow scaling and strength to match the observed
slope and amplitude.  In this model, the $M_*-Z$ relation evolves slowly
from $z=6\rightarrow 2$; an (admittedly uncertain) extrapolation to $z=0$
broadly matches local $M_*-Z$ observations.  Overall, the $M_*-Z$
relation provides critical constraints on galactic outflow processes
during the heydey of star formation in the Universe.
\end{abstract}
\maketitle
\section{Introduction}

Galactic outflows are observed to carry mass and metals into the
intergalactic medium~(e.g. \cite{vei05}).  At high redshift, outflows are
ubiquitous from star-forming galaxies~(\cite{pet01}), and the metals
that they carry are seen in a wide range of environments (\cite{pet06}).
\cite{pet99} originally noted the missing metals problem, namely that
the amount of metals present in Lyman break galaxies at $z\sim 2.5$
falls far short of the amount expected to be produced by their stars
(see \cite{dav06b} for an updated discussion).  Hence outflows appear to
have a large impact on the evolution of galaxies during the heydey of
star formation in the Universe, and metals provide a convenient avenue
with which to trace and quantify ejected gas.

In \cite{opp06} (hereafter OD06) we implemented a number of parameterized
models for galactic outflows into \gad\ cosmological hydrodynamic
simulations.  We tested these models against observations of \ion{C}{iv}
quasar absorption line systems at $z\sim 2-6$, and found that only a
relatively narrow range of outflow parameters were capable of enriching
the diffuse intergalactic medium (IGM) early enough while not overheating
it.  Intriguingly, the most successful models were ones that employed a
momentum-driven wind scenario~(\cite{mur05}) that is also favored by local
observations of starburst outflows~(\cite{mar05}).  Hence it is plausible
that distant outflow-driving galaxies obey scalings similar to local ones.

Another critical test of such outflow models is measuring the metals
that get left behind in galaxies rather than driven into the IGM.
Observationally, this can be traced using the galaxy mass-metallicity
relation.  \cite{tre04}, using the Sloan Digital Sky Survey, found that
low-redshift emission-line galaxies had a gas-phase metallicity that
increased with stellar mass up to $\sim 10^{10.5}M_\odot$, and then
flattened to higher masses.  At $z\sim 1$, smaller survey areas make
it difficult to track the high-mass flattening, but indications are
that galaxies at a given mass are only mildly underenriched compared
to present-day ones~(\cite{sav05}).  This evolution was extended out
to $z\sim 2$ by \cite{erb06}, who found that galaxies back then are
roughly one-half as metal-rich as galaxies today at a given stellar mass.
This is mildly surprising as only about one-quarter of all stars (and
therefore, presumably, metals) have formed by then~(\cite{rud03}).
However, most models of cosmic chemical evolution predict earlier
enrichment in the highly biased environments that form early stars
(e.g. \cite{cen99}, \cite{dav06}).  In any case, if the solution to the
missing metals problem is that the majority of metals have been ejected
from galaxies~(e.g. \cite{dav06b}), then the amount and distribution
of metals left behind in galaxies likely offers key insights into the
nature of outflows.

In these proceedings we test our outflow models in OD06 against the
$z\approx 2$ mass-metallicity data of \cite{erb06}, and study the
evolution of the mass-metallicity relation.  We show that the same
outflow model that best matches the IGM data also best matches the
mass-metallicity relation, and furthermore the models that do not match
the IGM data provide a poorer fit.  This lends further support to the
idea that momentum-driven galactic winds in a hierarchical galaxy formation
setting can explain a wide range of observables at $z\geq2$.

\section{Simulations}

We employ the suite of \gad\ cosmological hydrodynamic simulations
described in OD06.  Briefly, these use the entropy-conservative
PM-Tree-SPH code \gad~(\cite{spr03a}) with improvements as described
in OD06 including metal-line cooling.  An $\Omega=0.3$ $\Lambda$CDM
cosmology was assumed, and cubic random volumes of $16$ and $32\hmpc$
on a side were represented with $2\times 256^3$ particles (the $8\hmpc$
runs are not useful for the present study).  The minimum resolved galaxy
stellar masses in these runs, corresponding to 64 star particles, are
$1.24\times 10^8M_\odot$ and $9.9\times 10^8M_\odot$, respectively.  

For conciseness, we focus on three of the six outflow models in OD06:
``no winds" (nw), ``constant winds" (cw), and ``momentum-driven winds"
(vzw).  The no winds case is shown for illustrative purposes, the constant
winds case is the outflow model in the runs of \cite{spr03b}, and the
vzw momentum-driven winds case is the one that, overall, best matches
the IGM metal line observations.  The mzw momentum-driven model of OD06
gives very similar results to vzw, and vzw seems mildly more plausible
based on observations of local starburst outflows~(\cite{rup05}).  As a
reminder, vzw assumes $v_{\rm wind}\propto \sigma$ and mass loading
factor $\eta\propto \sigma^{-1}$ (where $\sigma$ is the galaxy velocity
dispersion), while cw assumes $v_{\rm wind}=484$~km/s and $\eta=2$ for
all galaxies.

To compare to the mass-metallicity relation observed in galaxies, we
calculate the star formation rate (SFR)-weighted metallicity of gas
particles in galaxies identified in our simulations.  This weighting is
intended to emulate how galaxy metallicities are typically measured, using
nebular emission lines that arise from warm ionized gas in star forming regions.
In \cite{erb06}, gas masses are obtained from star formation rates using
the Kennicutt relation together with the H$\alpha$ radius, whereas
in our simulations gas masses are obtained as bound cold, dense gas.
These measures are quite different and their relation is unclear, and
further there are possible systematics in both simulations and data, so
we caution that comparisons to gas mass relations are highly preliminary.
On the other hand, comparisons to the stellar mass-metallicity relation
should be reasonably robust, and that will be the main focus of
these proceedings.

\section{Outflows and Mass-Metallicity Relation}

\begin{figure}
\centering
\vskip -0.5in
\includegraphics[width=110mm]{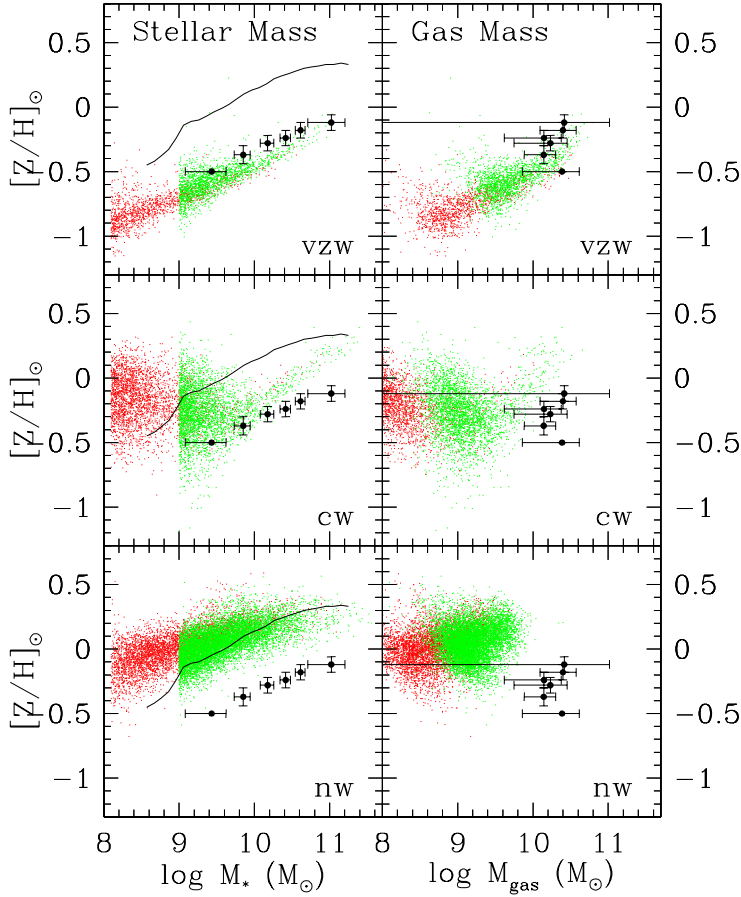}
\vskip -0.6in
\caption{Stellar mass-metallicity (left panels) and gas mass-metallicity
(right panels) relations from the vzw, cw, and nw runs (top to bottom panels)
at $z=2$.  Red and green points show $16\hmpc$ and $32\hmpc$ box galaxies,
respectively.
Data points shown are from \cite{erb06}; the thick solid line in the
left panels is the median $M_*-Z$ relation at $z\approx 0$ from \cite{tre04}.
\label{fig:metcomp}}
\vskip -0.1in
\end{figure}

Figure~\ref{fig:metcomp} shows galaxy metallicities versus stellar mass
(left panels) and gas mass (right panels), for the vzw, cw, and nw models
(top to bottom).  Overplotted are data points from \cite{erb06} and
the median stellar mass-metallicity relation in SDSS from \cite{tre04}
(solid line).  The $z\approx 0$ relation is shifted up by about $\times
2$ in metallicity at a given stellar mass but has a similar slope,
as noted in \cite{erb06}.

Figure~\ref{fig:metcomp} demonstrates, at its most basic level, that
the outflow model has a significant impact on the mass-metallicity
relation.  Therefore both the slope and the amplitude of this relation 
provide critical constraints on outflows.  If no outflows are included
(nw model), not only are too many stars produced (see e.g. OD06), but too
many of the metals remain locked into galaxies at a given stellar mass.
Hence not only are outflows required to enrich the IGM, they are concurrently
required to {\it de-}enrich galaxies.

Comparing our two outflow models reveals some interesting and perhaps
surprising insights into the nature of the mass-metallicity relation.
Given that the constant wind (cw) model (middle panels) assumes a constant
outflow velocity of $484$~km/s, one might expect a characteristic
feature in the $M_*-Z$ relation corresponding to halos of this escape
velocity ($\sim 2\times 10^{13}M_\odot$).  In particular, halos above
that mass should have a flat $M_*-Z$ relation (because they produce
metals in proportion to their stellar content), while below that mass
there should be a slope arising from the fact that outflows can escape
more easily from smaller halos.  Indeed, the overall shape of the local
$M_*-Z$ relation has been qualitatively explained using such a scenario.
However, at face value, the trend produced in the cw model is exactly
the opposite: At low masses, there is a constant $M_*-Z$ relation,
while at $M_*>10^{10}$ it begins to rise sharply.

Why does this happen?  The answer is not entirely clear.  A preliminary
idea that we are now investigating is that because the constant wind model
so widely enriches the IGM at an early epoch (see OD06, Figure~10), it
{\it pre-pollutes} the gas out of which the smaller galaxies later form.
In fact, the smallest galaxies forming most recently actually have
a slightly higher metallicity, i.e. there is an inversion in $M_*-Z$ at
low masses, because the IGM becomes more enriched overall with time.  This
suggests that environmental effects and clustering must also be taken
into account when interpreting mass-metallicity relations.  It is also
worth noting that, although the $z\approx 2$ data at $M<10^{10}M_\odot$
are too uncertain to definitively rule out such a flattening at low masses,
locally the $M_*-Z$ slope is seen to continue relatively unbroken to
quite small masses~(\cite{lee06}), which would clearly rule out such a
high level of pre-pollution.

Next, there is the issues of the slope of the $M_*-Z$ relation at
$M_*>10^{10}$; it is far too steep compared with observations, agreeing at
$M_*\sim 10^{10}$ but being a factor of two too high by $M_*\sim 10^{11}$.
Hence a constant outflow velocity actually produces an incorrect $M_*-Z$
slope by being over-efficient at expelling metals from small galaxies
as compared to large ones.  This favors a scenario where smaller
galaxies have smaller outflow velocities.  At some large mass, the cw
model is expected to produce a flattened $M_*-Z$ slope, but halos with
an escape velocity of $\sim 500$~km/s are rare within our volume by $z=2$
so it is not apparent here.

The vzw model (top left panel) agrees quite well with observations,
matching both the observed slope and the amplitude at $z\approx 2$.
The amplitude agreement reflects a proper balance between metals retained
in galaxies and expelled into the IGM, and is related to the fact that
vzw broadly reproduces the stellar mass density evolution.  The agreement
in the slope, more interestingly, may be indicating that {\it galaxies
lose a fixed fraction of their expelled material.}  This is precisely
the scenario in the vzw model because the outflow speed scales with
the escape velocity, and hence all halos (to first order) lose material
equally efficiently.  It also happens in the no-wind case, in the sense
that the fraction lost is essentially zero for all galaxies, and despite
its other failings it does reproduce the slope of the $M_*-Z$ relation.
Outflow models like cw, where small galaxies lose a higher fraction of
their expelled material, result in too steep a slope.  Interestingly,
\cite{erb06} finds that observationally there is no evidence that small
galaxies at $z\approx 2$ have preferentially lost more of their baryons
as compared to more massive ones.

Turning to the gas mass-metallicity relation (right panels), the
observations span a rather limited range in gas mass, and both cw and
vzw are in broad agreement with data given the level of uncertainties
discussed earlier.  In contrast, with no outflows the gas is too quickly
converted into stars, leaving no objects with gas masses as observed.
This shows that outflows are required in order to keep galaxies as
gas-rich as observed at these epochs.  The predicted trends in metallicity
versus gas mass and versus stellar mass are similar, which reflects the steady
conversion of gas into stars in simulations.

In summary, comparisons of various outflow models to the stellar
mass-metallicity relation of galaxies shows that outflows must be strong
enough to eject significant amounts of metals, but not be so strong
as to over-pollute the diffuse IGM out of which later galaxies form.
The slope of the mass-metallicity relationship suggests that galaxies
should lose a fixed fraction of their outflowing metals to the IGM,
favoring a scenario where the outflow speed scales as the escape velocity.
It is fairly remarkable that our momentum-driven wind model naturally
produces the correct strength and scaling of outflows in order to
reproduce the observed $M_*-Z$ relation.  It is even more remarkable
when one considers that this model is fairly uniquely succesful at enriching
the IGM to observed levels at these epochs, and suggests that such models
are on the right track towards understanding the basic scaling relations 
of outflows at all epochs.

\section{Mass-Metallicity Evolution}

\begin{figure}
\centering
\vskip -0.5in
\includegraphics[width=110mm]{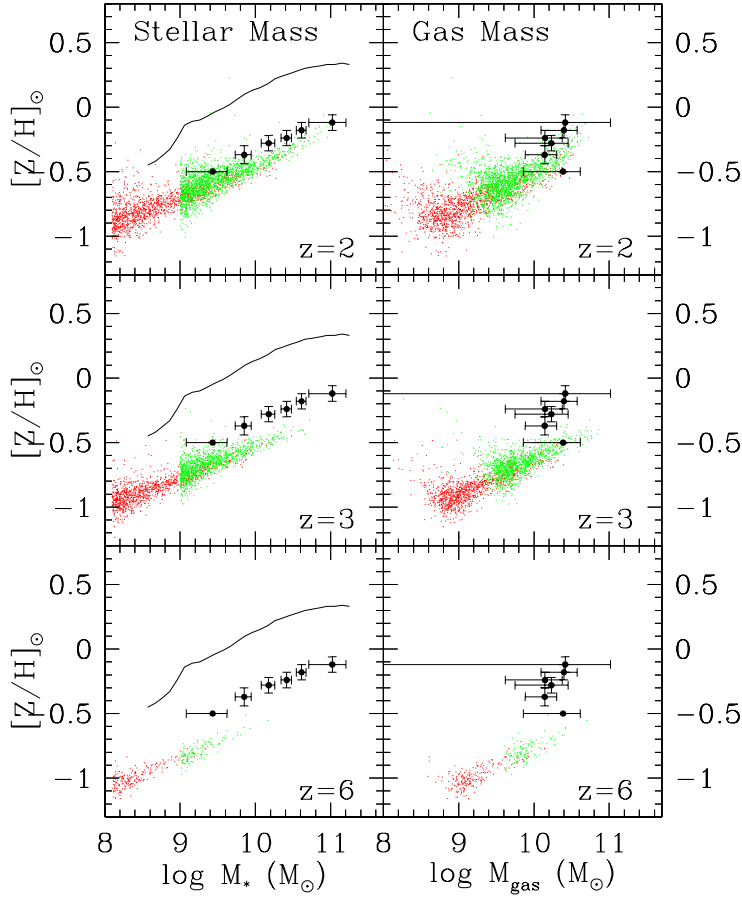}
\vskip -0.6in
\caption{Stellar mass-metallicity (left panels) and gas mass-metallicity
(right panels) relations from the vzw runs at $z=2,3,6$ (top to bottom).  
$16\hmpc$ box is shown
as red points, $32\hmpc$ box as green points ($>10^9M_\odot$). 
Data points shown are from \cite{erb06}; the thick solid line in the
left panels is the median $M_*-Z$ relation at $z\approx 0$ from \cite{tre04}.
\label{fig:MZevol}}
\vskip -0.1in
\end{figure}

Figure~\ref{fig:MZevol} shows galaxy metallicities versus stellar mass
(left panels) and gas mass (right panels), at $z=2,3,6$ (top to bottom)
from the vzw model.  The amplitude of the mass-metallicity relation in
this model evolves slowly, with galaxies already being substantially
enriched by $z=6$~(\cite{dav06}), while the slope is set early on and
remains esentially constant at $Z\propto M_*^{0.3}$ (in agreement with
the $z\approx0$ determination by \cite{lee06}).

A simple linear time-based extrapolation from $z=6\rightarrow 2\rightarrow
0$ indicates that this model would be very close to matching the
$z\approx 0$ data from \cite{tre04}.  The increase in amplitude is about
0.06~dex per Gyr from $z=6\rightarrow 2$.  At $z=0$ this extrapolation
would predict [Z/H]$_\odot=0.2$ at $M_*=10^{10}M_\odot$.  However,
a linear extrapolation may not be appropriate given that outflows
are likely less prevalant at low-$z$.  It is also worth putting in a
cautionary note about systematics in observed metallicity indicators;
commonly-used indicators may differ by $\sim\times 2$ from the true
gas-phase metallicity~(\cite{ell05}), so one must be cautious not to
over-interpret the precise values.

Overall, it is interesting that galaxies seem to move towards higher
masses predominantly {\it along} the $M_*-Z$ relation.  The slope of this
relation ($\approx 0.3$), does not appear to have a natural physical
explanation.  Although \cite{tre04} interpreted it in terms of a leaky
closed box model, such a scenario seems overly simplistic, as it does
not incorporate hierarchical growth and environmental effects, both
of which are important in our simulations for establishing the $M_*-Z$
relation.  Instead, the slope probably reflects some particular balance
between outflows and galaxy growth, further emphasizing that the slope,
amplitude, and evolution of the mass-metallicity relation all provide critical
tests of outflow models.

\section{Conclusions}

Using our cosmological hydrodynamic simulations incorporating galactic
outflows, we investigate the nature of the mass-metallicity relation in
galaxies at high redshift, and compare it to $z\approx 2$ observations
by \cite{erb06}.  We find that:
\begin{itemize}
\item Outflows are required in order to reduce the metallicity of
galaxies at $z=2$.  With no outflows, galaxies at a given stellar mass 
are twice as enriched as they should be by that epoch, comparable 
to galaxies today.
\item The slope and amplitude of the $M_*-Z$ relation is naturally reproduced 
in our momentum-driven wind model that is also favored from 
comparisons to IGM enrichment observations from $z\approx 5\rightarrow 2$.
\item The constant wind-speed outflow model produces an $M_*-Z$ relation 
that is flat at low masses, possibly owing to pre-pollution from early 
widespread winds, and rises too steeply at high masses, both of which
are in disagreement with observations. Hence it does not appear that
smaller galaxies preferentially lose more material to outflows at
these epochs.
\item The mass-metallicity slope is established at early times ($z>6$)
and remains constant at $Z\propto M_*^{0.3}$, while the amplitude evolves
slowly with redshift ($\approx 0.06$~dex per Gyr).  Galaxies tend to
evolve mostly along the $M_*-Z$ relation towards higher stellar masses.
\item A simple, albeit questionable, extrapolation to $z=0$ yields 
a relation that is in good agreement with data from~\cite{tre04}.
The unbroken slope to the lowest observed masses~(\cite{lee06}) provides
another non-trivial constraint that is met in our momentum-driven
wind scenario.
\end{itemize}

In summary, the mass-metallicity relation provides another critical
constraint on galactic outflows across cosmic time.  Understanding the
complex processes that establish the shape and evolution of the $M_*-Z$
relation will likely require moving beyond simple closed box model
variants, and may benefit from insights gained using numerical simulations that
include metal production and distribution mechanisms.  The comparisons
here strengthen our claim that momentum-driven wind scalings (whether or
not they actually arise from momentum-driven winds) are able to properly
distribute metals throughout the Universe at high redshfits.


\end{document}